\documentclass[10pt,twocolumn,twoside]{IEEEtran}
\usepackage{cite}
\usepackage[dvips]{graphicx}
\usepackage{amssymb}
\usepackage{amsmath}
\usepackage{esvect}

\usepackage{lipsum}
\usepackage{epstopdf}
\usepackage{color, colortbl, multirow}
\definecolor{Gray1}{gray}{0.9}

\usepackage{pdflscape, algorithm, algpseudocode, booktabs, hhline}

\newcommand\doubleRule{\toprule\toprule}

\ifCLASSINFOpdf

\else

\fi

\begin{document}
\title{Delineation and Analysis of Seismocardiographic Systole and Diastole Profiles}


\author{Tilendra Choudhary,~\IEEEmembership{Member,~IEEE,}
        M.K. Bhuyan,~\IEEEmembership{Senior Member,~IEEE},
        and~L.N. Sharma

\thanks{T. Choudhary, M.K. Bhuyan, and L.N. Sharma are with the Department of Electronics and Electrical Engineering, Indian Institute of Technology Guwahati, Assam-781039, India (e-mail: \{tilendra, mkb, lns\}@iitg.ac.in).}}



\maketitle

\begin{abstract}
Precise estimation of fiducial points of a seismocardiogram (SCG) signal is a challenging problem for its clinical usage. 
Delineation techniques proposed in the existing literature do not estimate all the clinically significant points of a SCG signal, simultaneously.  
The aim of this research work is to propose a delineation framework to identify IM, AO, IC, AC, pAC and MO fiducial points with the help of a PPG signal. 
The proposed delineation method processes a wavelet based scalographic PPG and an envelope construction scheme is proposed to estimate the prominent peaks of the PPG signal. A set of amplitude-histogram based decision rules is developed for estimation of SCG diastole phases, namely AC, pAC and MO.
Subsequently, the systolic phases, IM, AO and IC are detected by applying diastole masking on SCG and decision rules.
Experimental results on real-time SCG signals acquired from our designed data-acquisition-circuitry and their analysis show effectiveness of the proposed scheme. 
Additionally, these estimated parameters are analyzed to show the discrimination between normal breathing and breathlessness conditions.

\end{abstract}



\section{introduction}

\begin{figure}
\centering
\includegraphics[width=8cm]{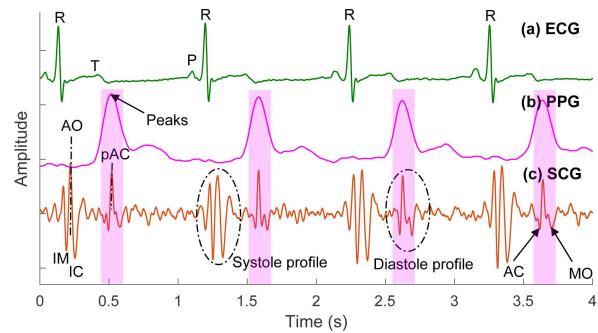}
\vspace{-0.1cm}
  \caption{Simultaneous recordings of ECG, fingertip-PPG and SCG signals.}
  \label{fig:intro}
  \vspace{-0.4cm}
\end{figure}

\IEEEPARstart{P}{hysiological} signal generated through heart mechanical actions, such as seismocardiogram (SCG) can be employed for assessment of cardiac performance \cite{rienzo2013_02, taebi2019}.
The SCG signal reflects cardiac-generated mechanical vibrations, which are recorded by mounting an accelerometer on the chest surface \cite{inan2015, sorensen2018nature}.
A heart cycle includes many important cardiovascular events, such as aortic valve opening (AO) and its closure (AC), mitral valve opening (MO) and its closure (MC),  rapid blood filling (RF) and rapid ejection (RE), isovolumic moment (IM), and isotonic contraction (IC).
Peaks and troughs of systole and diastole profiles of a SCG cycle indicate these events, which are called fiducial points \cite{choudhary2018JBHI}. 
Delineation of SCG waveform refers to the determination of fiducial points and the estimation of time intervals between the fiducial points, which are clinically significant for pathological interpretations of a cardiovascular system \cite{sorensen2018nature}.
The current researches mainly focus on the definition and identification of fiducial points of a SCG signal and relate them with the echocardiogram for identification of physiological events of a cardiac cycle \cite{crow1994, sorensen2018nature}. 
Fig. \ref{fig:intro} illustrates the simultaneous recordings of ECG, fingertip-photoplethysmogram (PPG), and SCG signals. 
Rhythmic variations of blood-volume at different parts of the body is generally observed due to the pumping actions of the heart, which is indicated by a PPG signal \cite{hertzman1938}.

The delineation of a SCG signal helps in evaluation of cardiovascular parameters, such as extraction and analysis of heartbeats, estimation of heart rate variability (HRV)  \cite{khavar2016, khavar2015_02, nguyen2012, tadi2015, khavar2015_01, shafiq2016, xu1996_01, choudhary2019BSPC, laurin2016}, hemodynamic parameter estimation \cite{tavakolian2010}, association with other cardiac signals \cite{choudhary2018SCGPCG}, detection of valvular defects and diagnosis of other cardiovascular diseases \cite{paukkunen2016, salerno1991_01}. 
The delineation can also improve the ability of wearable SCG sensor nodes in the development of pervasive wireless body area networks \cite{rienzo2013_02}.
An extensive investigation has been done for annotation of AO phase in the SCG signal \cite{choudhary2018JBHI, khavar2016, khavar2015_02, nguyen2012, tadi2015, khavar2015_01, shafiq2016, choudhary2019BSPC, laurin2016, choudhary2019ISJ}. 
Most of the studies employed ECG as a reference signal for finding out cardiac information from a SCG signal \cite{tadi2015, shafiq2016, choudhary2019BSPC}.
The determination of IM, AO and AC points are proposed in \cite{khavar2015_02} using a high pass filter and a triple integration technique. 
To annotate AC and IM points, high frequency cardiac acceleration (HFACC) envelope was suggested in \cite{khavar2015_01}.
Khavar {\it et al.} presented a method to estimate IM, AO and AC phases in SCG signal \cite{khavar2016}.
 The method employed a high pass filter, a moving averaging filter, an extrema search block and a probabilistic decision making technique on systolic, diastolic and heart-rate envelopes. 
In \cite{mora2019}, the SCG systole points are annotated using calibration, heartbeat detection, and prototype matching scheme.
By utilizing orthogonal subspace projection scheme, the fiducial points AO and pAC (peak just after AC) were estimated in our previous reported work \cite{choudhary2019BSPC}.

\begin{figure}
\centering
\includegraphics[height=3cm, width=8.4cm]{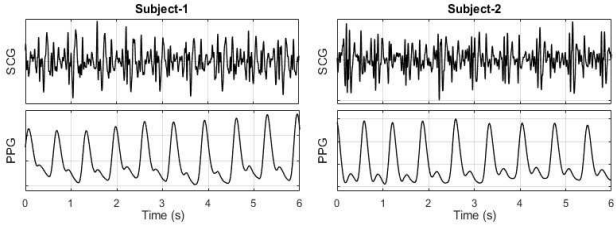}
\vspace{-0.3cm}
  \caption{Difficulty in localization of a SCG profile in presence of interbeat and intrabeat variabilities in contrast to PPG signals.}
  \label{fig:motivation}
  \vspace{-0.4cm}
\end{figure}

Typically, the AO point creates a prominent peak under the systolic profile of the SCG signal \cite{choudhary2019ISJ}.
Thus, other informative systolic points may be estimated with the help of the AO peaks. 
However, delineation of a SCG diastole profile is still a challenge, and the task is even more complicated when both systole and diastole profiles have similar characteristics. It is observed in Fig. \ref{fig:intro} that the diastole delineation would be comparatively easier with the help of a PPG signal.
This is due to the fact that projections of PPG peaks lie on AC-to-MO segment of a SCG signal. The PPG signal is less noisy, while a SCG signal changes its basic characteristics if the sensor placement is affected by the body movements. 
As illustrated by a concurrently recorded SCG and PPG signals from two subjects in Fig. \ref{fig:motivation}, it is difficult to localize a SCG profile due to inter and intra beat variabilities. 
Moreover, the existing methods are limited to estimate only a few fiducial points, through which the characterization of both systolic and diastolic profiles of the SCG would not be possible.




To address these issues, a PPG-assisted automatic delineation framework for a SCG signal is proposed, which can estimate IM, AO, IC, AC, pAC and MO fiducial points in systole and diastole profiles. The framework can also estimate two important cardiac intervals, namely IVRT (AC-to-MO) and LVET (AO-to-AC). 
Multiscale scalogram is employed for spectro-temporal localization of nonstationary and nonlinear signals \cite{bialasiewicz2013}.
To analyze the SCG waveform, a PPG signal is employed and wavelet scalogram-based time-frequency analysis is performed onto it. 
An envelope construction method is proposed and decision rules are used for estimation of fiducial points.
In addition, one of its applications is presented in cardiorespiratory analysis in order to show the ability of the delineated points.
Since a change in the respiratory effort-level affects the morphology of the SCG waveform \cite{taebi2019}, \cite{choudhary2019TENSYMP}, the fiducial points of a SCG signal may be used in assessing the fitness of a respiratory system.
As an extension of the proposed delineation work, the breathlessness (shortness of breath or dyspnea) and normal breathing conditions are perfectly discriminated by the delineated points.
The proposed method is elaborately discussed in the following sections. 
 


\section{Proposed Method}
Wavelet transform (WT) represents a signal as linear combinations of a basis function-set and its variations. The variation is achieved by means of scaling ($l$) and shifting ($m$) parameters of a single mother wavelet function $\psi(t)$. So, signal $x(t)$ can be analyzed using WT as \cite{hussein2015}:
\begin{equation}
w_{l,m}\{x\}=\dfrac{1}{\sqrt{a}} \int_{-\infty}^{+\infty} x(t) \psi^{*}\left(\dfrac{t-m}{l}\right) dt, ~l>0
\end{equation}

\noindent
where, $w_{l,m}$ represents $m^{\text{th}}$ wavelet coefficient at $l^{\text{th}}$ scale and the asterisk (${*}$) is a complex conjugate operator.  
 Since, the second derivative of a Gaussian function ({\it Gaus-2}) has a morphological similarity with the pulsatile profile of a PPG signal, it is chosen as the mother wavelet for continuous wavelet transform (CWT).
The prototype wavelet  $\psi(t)$ can be defined as \cite{addison2005}:
\begin{equation}
\psi(t)=-\dfrac{dg(t)}{dt}=\text{C}(1-t^2)e^{-t^2/2}
\end{equation}
\noindent
where, $g(t)$ and $\text{C}$ denote a Gaussian function $\mathcal{N}(0,\,1)$ and normalization factor, respectively. 

\begin{figure*}
\centering
\includegraphics[height=8cm, width=16.5cm]{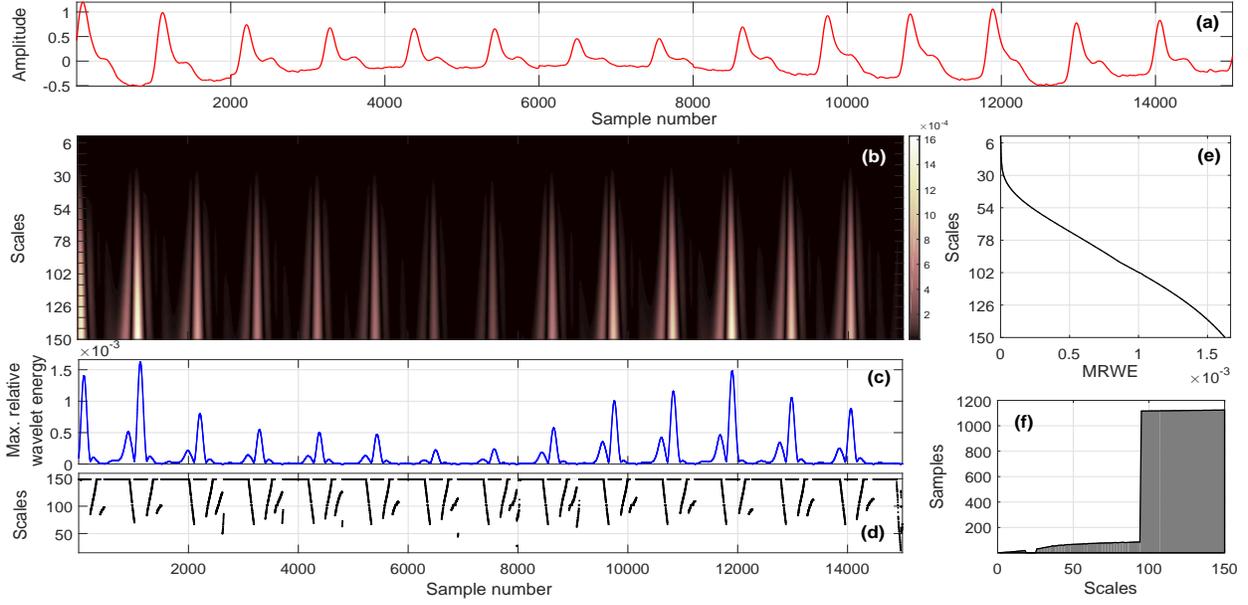}\\
  \caption{Scalogram of PPG signal. (a) Input PPG signal, (b) PPG scalogram, (c) change of maximum relative wavelet energy (MRWE) with time, (d) variation of scales with time, (e) predominating energy (MRWE) with the higher scales, and (f) overlapping area of time and scales.}
  \label{fig:scalogram}
  \vspace{-0.2cm}
\end{figure*}

\subsection{Processing of Scalogram}
A scalogram shows 3-D multiscale distributions of signal components with the help of wavelet transformation \cite{bialasiewicz2013}, \cite{addison2005}.   
It indicates the relative wavelet energy for each coefficient $w_{l,m}$, which is obtained as \cite{addison2005}:
\begin{equation}
E_{l,m} = |w_{l,m}|^2
\end{equation}
\begin{equation}
S_{l,m}   \longleftarrow \frac{E_{l,m}}{\sum_l \sum_m E_{l,m}} 
\end{equation}

\noindent
where, $E_{l,m}$ and $S_{l,m}$ denote wavelet energy density and scalogram, respectively.
A PPG-scalogram is obtained by processing an input PPG signal upto 150 scales, which is illustrated in Fig.~\ref{fig:scalogram}. 
The corresponding scalogram is shown in Fig.~\ref{fig:scalogram}(b) and subsequently, its all possible two dimensional interpretations are shown in Fig.~\ref{fig:scalogram}(c)--(f). 
From this representation, a time series of maximum relative wavelet energy (MRWE) is obtained irrespective to the scales as shown in Fig.~\ref{fig:scalogram}(c). 
A larger MRWE value corresponds maximum resemblance between the wavelet function and the pulsatile waveform, which in turn gives an indication of the presence of PPG peaks. 
Thus, the MRWE signal is selected for further analysis.
The PPG signal is detrended and subsequently, added up with the MRWE signal to emphasize the PPG-peaks and to suppress other spurious peaks.
The detrending process removes the baseline-drift (BD) from the signal, which uses a high pass filtering scheme. 
The signal produced by ensembled-sum is denoted by $x_p^*[n]$.

\begin{figure}
\centering
\includegraphics[height=5.5cm]{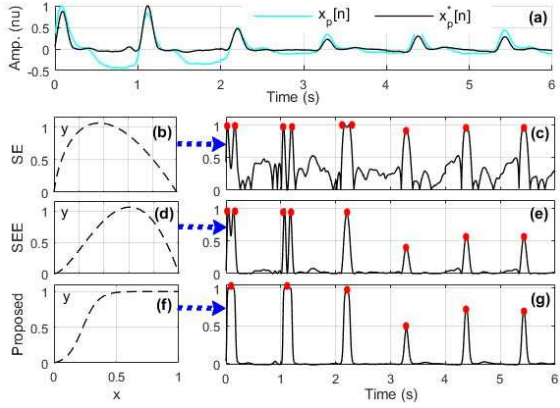}
\vspace{-0.2cm}
  \caption{Performance comparison of the proposed envelope construction for $x_p^*[n]$. (a) PPG and ensemble sum signal; (b)-(g) transfer characteristics and envelope profile estimation using using SE, SEE and the proposed method.
   Red dots denote local peaks of the envelope impulses.}
  \label{fig:env}
\end{figure}

\begin{figure}
\hbox{\hspace{-0.3cm}
\includegraphics[height=4cm]{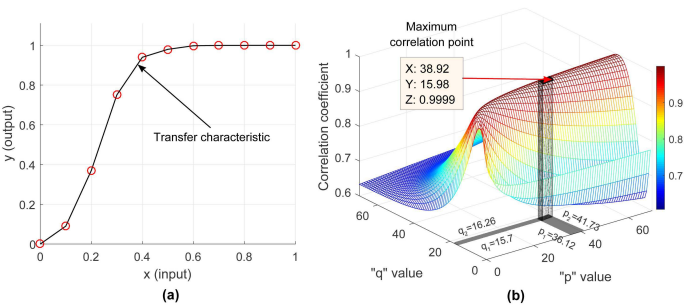}}
  \caption{Estimation of modeling parameters. (a) Desired transfer characteristic in a piecewise linear fashion, and (b) mesh plot representing variations of correlation for obtained characteristic corresponding to curve fitting parameters `$p$' and `$q$' [Eq. (\ref{Eq:y_eq})]. Estimated parameters corresponding to maximum correlation are highlighted with 95\% confidence interval using gray color bands. Transfer function produced by fitted parameters achieves maximum correlation.}
  \label{fig:modeling}
  \vspace{-0.2cm}
\end{figure}

\subsection{Proposed Envelope Construction Scheme}

The Shannon entropy (SE) and Shannon energy (SEE) based envelopes produce minima point for higher normalized-signal-amplitude. 
The methods for estimating envelopes such as absolute, energy, analytic signal, SE and SEE were compared in \cite{choudhary2018JBHI}.
The ensembled-sum signal $x_p^*[n]$ is employed for envelope construction using SE and SEE as shown in Fig. \ref{fig:env}(b)--(e). They provide bifurcated profiles for input peaks having large relative amplitudes. 
Hence, a modified transfer characteristic is proposed as shown in Fig.~\ref{fig:env}(f), which can effectively estimate the envelope profile of  $x_p^*[n]$ by creating sharp impulses at the locations of PPG-peaks (Fig.~\ref{fig:env}(g)).
In contrast to other methods, it emphasizes the PPG-peaks without bifurcation and suppresses the other regions relatively better. 
The proposed profile can be obtained via fitting an exponential curve in a least square sense.  
Mathematically, it can be represented via two exponential expressions, $(1-e^{-qx})$ and $1/(1+e^{-qx})$, where $x$ corresponds to the input vector. 
Hence, the function for fitting of the desired characteristic can be expressed as: 
\begin{eqnarray}
y&=&\dfrac{1-e^{-qx}}{1+pe^{-qx}}, ~~~ \forall x\in [0, 1], ~~~p,q \in \mathbb{N} 
\label{Eq:y_eq}
\end{eqnarray}    


\noindent
To achieve the desired profile, the modelling parameters $p$ and $q$ need to be estimated for different input values.
Fig.~\ref{fig:modeling} shows a variation of correlation between desired and estimated characteristic. The estimated transfer characteristic is achieved by varying the parameters. 
The fitted parameters are also highlighted with 95\% confidence intervals, which are presented by gray rectangular patches in $p$-$q$ plane. By the usage of fitted $p$ and $q$ parameters, the designed model shows the highest correlation with the desired characteristic as shown in Fig.~\ref{fig:modeling}(b). 
Thus, in an average least square sense, parameters $p$ and $q$ are estimated as 39 and 16, respectively.
In the proposed method, this envelope extraction scheme is used for the detection of true PPG-peaks and later, for the estimation of AO peaks.

\subsection{Detection of PPG-Peaks}
This developed model is applied to the ensembled-sum signal $x_p^*[n]$ to get the desired envelope profile. The envelope provides localization of PPG-peaks in the form of sharp impulses. The peaks of these impulses can be easily identified using simple amplitude-temporal based thresholding schemes. 

\subsection{Delineation of Diastolic Regions}
The diastole profiles of a SCG signal are estimated with the help of PPG signals at the initial stage. 
The detected PPG-peaks are found useful in estimating the pAC peaks in SCG-diastole profile.  
The pAC point corresponds to the maxima in a SCG diastole, and it can  be located by searching the peaks of a PPG signal with the help of a narrow and symmetric window.
Subsequently, a proposed set of amplitude-histogram based decision rules can be employed for estimation of adjacent fiducial points (AC and MO) as follows: 

\subsubsection{Amplitude-Histogram based Proposed Decision Rules}
The decision rules (Algorithm~\ref{alg:code1}) block is employed on both diastole and systole profiles of a SCG signal.
At first, each of the SCG diastolic profiles is segmented into two small blocks adjacent to pAC peaks, and all local peaks are detected for both the blocks.
It is observed that amplitude of the peaks gradually decreases from pAC locations.
Hence, each of the blocks is employed for the following set of operations. 
Five frequency bins are created to show the distribution of local peaks with respect to the pAC point under a segmented block of diastole profile as listed in step~5~(Algorithm~\ref{alg:code1}).
Subsequently, all the frequency bins are scanned sequentially ($F^1$--$F^5$) to find non-zero elements (local peaks).
In the scanning process, the first bin corresponding to have non-zero elements is observed and selected for further investigation.
Among all these local peaks from the selected bin, the first significant peak closest to pAC is estimated, say $P_1$.
Then, the segment $P_1$ to pAC is applied for minima search for both the blocks, which can estimate AC and MO fiducial points.

\begin{algorithm}[!t]
\footnotesize
\caption{Proposed decision rules}
\label{alg:code1}
\begin{algorithmic}[1]
\renewcommand{\algorithmicrequire}{\textbf{Inputs:}}
\renewcommand{\algorithmicensure}{\textbf{Outputs:}}
\Require $x_s[n]$, ${pks}_{i=1,2,...,\#{pks}}$
\Statex // $x_s[n]$ corresponds to the input SCG signal
\Statex // $pks$ corresponds to pAC peaks in diastole (or AO peaks in systole analysis)
\State Segment the SCG profiles into two blocks: 
\Statex  blockA$_i = x_s[({pks}_i - 100 \text{ ms}) : {pks}_i]$
\Statex  blockB$_i = x_s[{pks}_i : ({pks}_i + 200 \text{ ms})]$
\State Detect all local peaks of blockA and blockB.
\State Process the following steps for each of the blocks:
\For{$i = 1$ to $\#{pks}$} \do \\           
  \State \textbf{(a)} Compute amplitude histogram of local peaks:                 
   \begin{equation}\nonumber 
   \rotatebox{90}{\hbox{\hspace{-0.8cm} frequency bins}}
\left \{\begin{aligned}
    & F_i^1   \gets   \text{peaks lying in amplitude range }(80-100\%] \text{ of } {pks}_i \\
    & F_i^2   \gets   \text{peaks lying in amplitude range }(60-80\%] \text{ of } {pks}_i \\
    & \vdots  \qquad \qquad \vdots \qquad \qquad \vdots \qquad \qquad \vdots \qquad \qquad \vdots \\
    & F_i^5   \gets   \text{peaks lying in amplitude range }(0-20\%] \text{ of } {pks}_i
  \end{aligned} \right.
\end{equation} 
  \State \textbf{(b)} Estimate first significant peak nearest to ${pks}_i$
          	\For{$j = 1$ to $5$} \do \\               
               \If{$n(F_i^j) \geq 1$} 
                 \Statex \qquad \qquad // where $n(\cdot)$ denotes number of elements.
              \State $P_{1i}~\gets$ Peak nearest to ${pks}_i$ in frequency bin $F_i^j$;
              \State break \textbf{for loop};
            \EndIf
          \EndFor
  \State \textbf{(c)} Detect minimum points between $P_{1i}$ and ${pks}_i$ from both 
      \Statex ~~~~~~~~~blocks A and B $\rightarrow$ 
  $M_{1i}$ and $M_{2i}$, respectively

\EndFor
 \Ensure Fiducial points: $M_1$, $M_2$ and ${pks}$
\end{algorithmic}
\end{algorithm}

\subsection{Delineation of Systolic Regions}
Once the fiducial parameters of the diastolic region are determined, delineation is carried out for the systole.
If the fiducial points, AC, pAC and MO are perfectly determined, then it is possible to precisely localize diastolic profiles in the SCG signal. 
 Usually, both systolic and diastolic profiles share the same spectral range and/or have similar amplitude strengths.
Due to that, identifying systolic profile in the presence of diastolic profile in a SCG cycle becomes a great challenge. 
Thus, the diastole profiles are masked in the SCG signal by replacing their corresponding samples with zero amplitude levels.
Typically, the AO peak possesses the prominent amplitude under a systole profile. But, its actual location cannot always be found by determining the maxima of the masked-SCG-cycle, which is created by a segment between any two same diastole-fiducials. 
The reason behind is sometimes, the AO peak has a smaller amplitude than MC or RE, which are the peaks adjacent to AO in SCG signal. 
The SCG segment, IM-AO-IC shows frequency range typically higher than other wave-segments in the SCG-systole.
So, the masked-SCG is bandpass filtered in a relatively higher frequency range of 20--30 Hz.
Due to that, the AO peaks are emphasized in the resulting signal.
Then, its upper envelope is extracted and subsequently, its instantaneous energy signal is employed to the proposed envelope construction technique (Eq. (\ref{Eq:y_eq})). 
Finally, the locations of all the AO peaks, which are indicated through impulses in the resultant signal, are detected using simple thresholding and peak correction scheme.
As like diastole profile, the proposed decision rules (Algorithm \ref{alg:code1}) are subsequently applied to the masked-SCG signal with detected AO peaks. It can localize two another systole fiducial points, IM and IC.

\begin{figure}
\includegraphics[height=3.9cm]{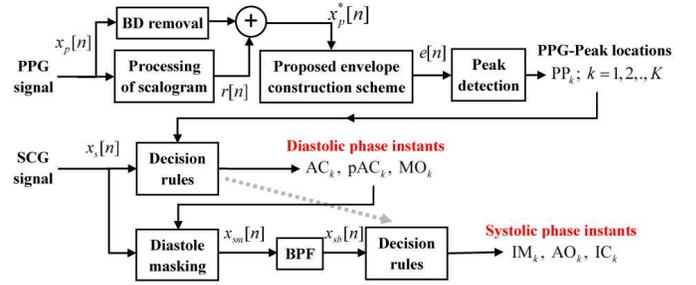}
  \caption{Block diagram of the proposed method.}
  \label{fig:block_dia}
  \vspace{-0.3cm}
\end{figure}

A block diagram of the proposed method for estimation of cardiac phases is shown in Fig. \ref{fig:block_dia}.
Initially, MRWE signal is extracted from the PPG-scalogram, which is further summed-up synchronously with the BD-suppressed PPG signal. Subsequently, an envelope is constructed on the output signal using the proposed scheme, and PPG-peaks are detected.
The PPG-peaks can detect the pAC peaks in the SCG signal.
Other essential diastole phase instants, AC and MO are estimated using the proposed decision rules.
After this process, the diastole profiles are localized and masked in the SCG signal, and subsequently, bandpass filtering is employed to estimate the AO instants. 
Similarly, other important systole phase instants, IM and IC are estimated using the proposed decision rules. 
Two important cardiac time intervals, namely LVET and IVRT can also be determined using the estimated AO, AC and MO parameters.

\begin{figure}[t!]
\centering
\includegraphics[height=7cm]{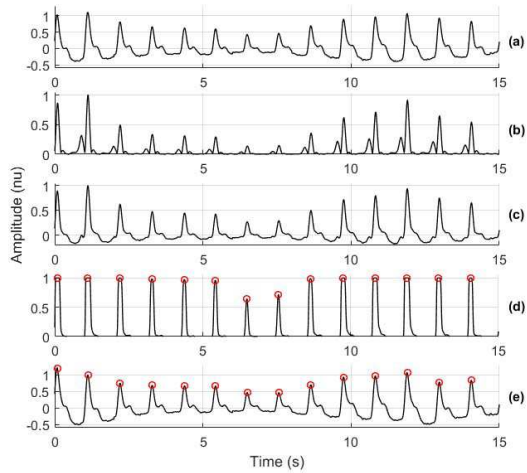}
\vspace{-0.6cm}
  \caption{Peak detection process of a PPG signal. (a) Detrended PPG $d[n]$, (b) MRWE signal $r[n]$, (c) $d[n]+r[n]$, (d) envelope, (e) peaks of the PPG.}
  \label{fig:ppg_peaks}
\end{figure}

\begin{figure}[t!]
\hbox{\hspace{0.11cm}
\includegraphics[height=4.5cm, width=9cm]{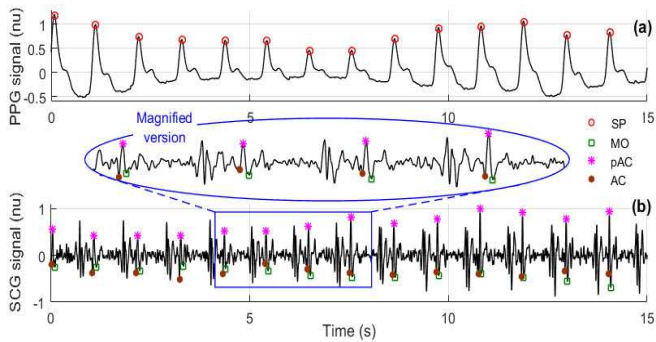}}
  \caption{Determination of parameters of SCG diastoles with the help of PPG peaks. (a) PPG signal with its peaks, (b) SCG signal with detected fiducial points (AC, pAC and MO) of diastoles using the proposed decision rules.}
  \label{fig:DiastolePoints}
\end{figure}

\begin{figure}[t]
\includegraphics[width=8.8cm, height=8cm]{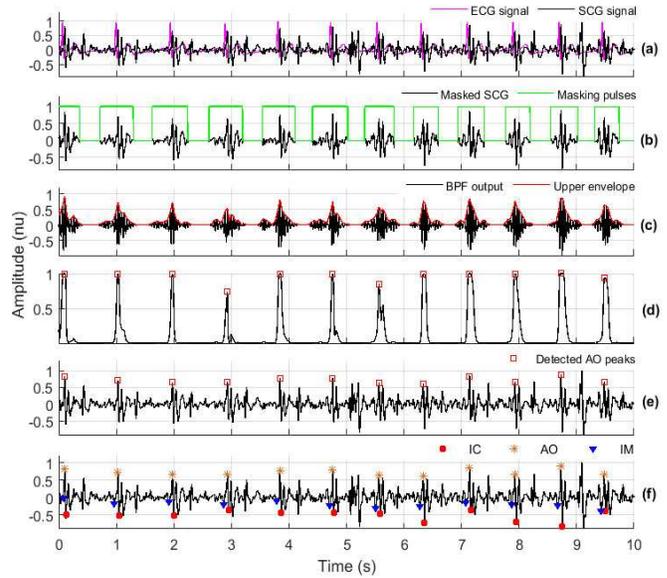}
\vspace{-0.3cm}
  \caption{Delineation process for systolic profiles of SCG. (a) Input SCG signal, (b) masking of diastole boundaries with the help of estimated diastolic fiducial points, (c) band pass filtered signal and its upper envelope construction, (d) final envelope created by using the proposed envelope scheme, (e) detected AO instants, and (f) estimated IM and IC instants using the decision rules.}
  \label{fig:SystolePoints1}
\end{figure}



\section{Experimental Setup}

\begin{table}[t!]
  \centering
  \caption{Demographics of subjects and their recording information}
   \scriptsize
  \newcommand{\m}{\hphantom{$-$}}
\newcommand{\cc}[1]{\multicolumn{1}{c}{#1}}
\renewcommand{\tabcolsep}{0.17pc}
\renewcommand{\arraystretch}{1.1}
    \begin{tabular}{ccccccccc}
    \doubleRule
    \textbf{\#Signals} & \textbf{\#Subjects} & \textbf{Sex} & \textbf{Age(y)} & \textbf{Wt.(kg)} & \textbf{Ht.} & \textbf{HR(bpm)} & \textbf{\#Systoles} & \textbf{\#Diastoles}  \\
    \midrule
   \multirow{2}{*}{16} & \multirow{2}{*}{8} & \multirow{2}{*}{M} & 28.75 & 71.63 & 5'7.6" & 79.18 & \multirow{2}{*}{3429} & \multirow{2}{*}{3425}  \\
      &    &       & $\pm$2.31  & $\pm$7.85  & $\pm$2.6"  & $\pm$10.93 &       &     \\

    \bottomrule
    \end{tabular}%
     \scriptsize
    \vspace{0.1cm}
   HR corresponds heart-rate, which is expressed in terms of beats per minute (bpm).
  \label{tab:demographics}%
  \vspace{-0.4cm}
\end{table}%

\subsection{Measurement Protocol and Data Acquisition}
\begin{figure}[t]
\includegraphics[height=4.5cm]{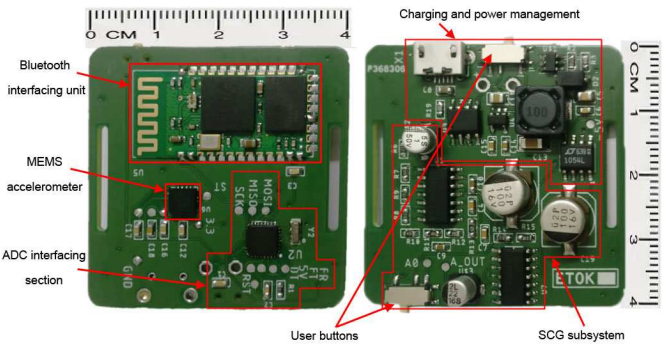}
\caption{Back and front views of our designed PCB for SCG signal recording.}
\label{fig:REC}
\end{figure}

\begin{figure}[t!]
\centering
\includegraphics[height=3cm]{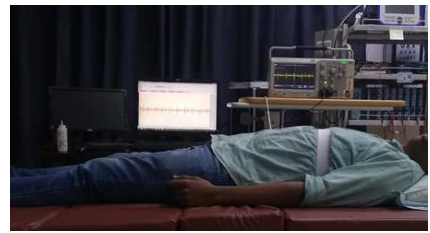}
\vspace{-0.2cm}
\caption{Experimental setup for recording of signals.}
\label{fig:exp_setup}
\vspace{-0.28cm}
\end{figure}

The SCG signals were acquired from eight male subjects by mounting the sensor node near lower end of the sternum on the chest wall. The demographics of involved subjects and the recording information are tabulated in Table \ref{tab:demographics}.
 Additionally, the ECG (Lead-II) and the PPG (at fingertip) signals were also acquired in a standard setting procedure.
All these three signals are digitized and synchronized using Biopac MP150 DAQ system at a sampling rate of 1 kHz. 
The signals were recorded in two different sessions: normal breathing for 5~minutes followed by holding breath for 50~s. 
So, breathlessness condition is artificially achieved for every subject by holding or stopping his respiratory activity. 
With an ethical approval and proper consent of subjects, the SCG signals were  recorded using a small sized signal recording system. 
The designed PCB for SCG signal acquisition is shown in Fig. \ref{fig:REC}.
The system consists of a miniaturized MEMS accelerometer (ADXL335, $\pm$3 g), a pre-amplifier, a Butterworth LPF (50 Hz cutoff frequency), buffer, data acquisition system (Biopac MP150), and a PC with the AcqKnowledge interfacing software. 
The experimental set-up for signal recording is shown in Fig. \ref{fig:exp_setup}.


\subsection{Experiment}
The proposed method is tested and validated with 16 different signals. 
According to breathing pattern, entire data can be divided into normal breathing (NBDB) and stopped breathing (SBDB) datasets. 
The proposed method employs simultaneously recorded SCG and PPG signals, while ECG is used as a reference signal for manual annotation of the SCG signal.
The proposed delineation framework is employed on each of the signals with a 10 s time window.
The PPG-peaks are determined initially with the help of wavelet scalogram, as shown in Fig. \ref{fig:ppg_peaks}, which further helps in locating the pAC-peaks under diastole profiles. Subsequently, the adjacent fiducial points AC and MO are estimated using the proposed histogram based decision rules. It is illustrated with an example in Fig. \ref{fig:DiastolePoints}.
The localization of diastolic region helps in precise detection of AO peaks under systole profiles, and the neighborhood fiducial points (IM and IC) can also be located with the help of same decision rules.  
The entire process of estimation of systolic parameters is shown in Fig.~\ref{fig:SystolePoints1}.
For both the profiles in the SCG waveform, the proposed envelope construction scheme helps in indicating the prominent peak locations.

\begin{figure}[t]\centering
\includegraphics[height=4cm]{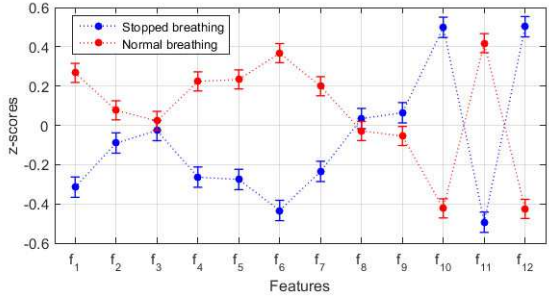}
\caption{Error-bar graph provides distribution of extracted features for both the classes. The distribution is shown via statistical parameters mean (red/blue color dot) and standard deviation (symmetric vertical lines around mean).}
\label{fig:Mean_Features}
\vspace{-0.2cm}
\end{figure}

\section{Results and Discussion}%
\subsection{Performance Evaluation}
The performance of the proposed delineation framework is analyzed for all the estimated parameters of both systolic and diastolic profiles.
The proposed method is tested with SCG signals of NBDB and SBDB databases. 
The performance is evaluated using three quantitative measures namely,  the sensitivity $\left( \text{Se}=\frac{\text{TP}}{\text{TP+FN}} \right)$, the positive predictivity $\left( \text{+P}=\frac{\text{TP}}{\text{TP+FP}}\right)$ and the accuracy $\left(\text{Acc}=\frac{\text{TP}}{\text{TP+FP+FN}}\right)$, where TP, FP and FN have their usual meaning \cite{wang2011ECG}.
The overall performance in the estimation of systolic fiducial points AO, IM and IC is summarized in Table \ref{tab:DetectResults_systole}.
The proposed method achieves average accuracies (in \%) of 95.2, 88.18 and 85.32 for AO, IM and IC points, respectively, on NBDB database, while 90.08, 86.09 and 89.13 are achieved for AO, IM and IC, respectively, on SBDB database.
The maximum sensitivity and positive predictivity are found in the detection of prominent AO peaks as compared to other systolic points. 

In a similar manner, the performance in the estimation of diastolic fiducial points pAC, AC and MO is summarized in Table \ref{tab:DetectResults_diastole}.
The proposed method achieves mean accuracies (in~\%) of 98.96, 97.55 and 94.54 for pAC, AC and MO points, respectively, on NBDB database, while 98.27, 95.06 and 95.06 are achieved for pAC, AC and MO, respectively, on SBDB database.
The maximum sensitivity and positive predictivity are found in the detection of pAC instants as compared to other diastolic points, which are 99.04 and 99.93 for NBDB, respectively, and 99.58 and 98.69 for SBDB, respectively. 
The obtained results clearly show the excellent performance of our delineation method.

\begin{table}
  \caption{Performance of the proposed method in the detection of systolic fiducial points}
   \scriptsize
  \newcommand{\m}{\hphantom{$-$}}
\newcommand{\cc}[1]{\multicolumn{1}{c}{#1}}
\renewcommand{\tabcolsep}{0.17pc}
\renewcommand{\arraystretch}{1.1}
    \begin{tabular}{ccccccccccc}
    \doubleRule
     & \multirow{2}{*}{\textbf{Record}} & \multicolumn{3}{c}{\textbf{AO Instant Detection}} & \multicolumn{3}{c}{\textbf{IM Instant Detection}} & \multicolumn{3}{c}{\textbf{IC Instant Detection}} \\
    \cmidrule(lr){3-5}   \cmidrule(lr){6-8}   \cmidrule(lr){9-11}
          &       & \textbf{Se}(\%) & \textbf{+P}(\%) & \textbf{Acc}(\%) & \textbf{Se}(\%) & \textbf{+P}(\%) & \textbf{Acc}(\%) & \textbf{Se}(\%) & \textbf{+P}(\%) & \textbf{Acc}(\%) \\
          \midrule
    \multirow{10}{*}{{\rotatebox[origin=c]{90}{\textbf{NBDB}}}} & s01a  & 99.11 & 99.7 & 98.81 & 96.73 & 97.31 & 94.2 & 99.1 & 99.4 & 98.52 \\
          & s02a  & 90.98 & 91.48 & 83.88 & 90.98 & 91.48 & 83.88 & 90.98 & 91.48 & 83.88 \\
          & s03a  & 98.28 & 99.31 & 97.61 & 79.11 & 79.93 & 66 & 85.27 & 86.16 & 75 \\
          & s04a  & 97.65 & 98.31 & 96.04 & 96.31 & 96.96 & 93.49 & 88.26 & 88.85 & 79.46 \\
          & s05a  & 95.51 & 95.73 & 91.61 & 88.42 & 89.26 & 79.91 & 94.56 & 94.79 & 89.89 \\
          & s06a  & 99.54 & 98.87 & 98.42 & 97.27 & 96.61 & 94.05 & 99.77 & 99.09 & 98.87 \\
          & s07a  & 98.91 & 99.56 & 98.47 & 98.47 & 99.12 & 97.61 & 75.49 & 75.99 & 60.95 \\
          & s08a  & 97.31 & 99.45 & 96.79 & 97.04 & 99.18 & 96.27 & 96.51 & 99.45 & 95.99 \\  
          \hhline{~----------}  \noalign{\vskip.2pt}
          & \multirow{2}{*}{\textbf{Overall}} & \textbf{97.16} & \textbf{97.8} & \textbf{95.2} & \textbf{93.04} & \textbf{93.73} & \textbf{88.18} & \textbf{91.24} & \textbf{91.9} & \textbf{85.32} \\ 
          &       & \textbf{$\pm$2.8} & \textbf{$\pm$2.86} & \textbf{$\pm$5.13} & \textbf{$\pm$6.63} & \textbf{$\pm$6.61} & \textbf{$\pm$10.9} & \textbf{$\pm$8.16} & \textbf{$\pm$8.18} & \textbf{$\pm$13.25} \\ 
          \hhline{-----------}  \noalign{\vskip.9pt}
          \rowcolor{Gray1}
   & s01b  & 100   & 100   & 100   & 100   & 100   & 100   & 100   & 100   & 100 \\  \rowcolor{Gray1}
          & s02b  & 75.41 & 86.79 & 67.65 & 75.41 & 86.79 & 67.65 & 75.41 & 86.79 & 67.65 \\  \rowcolor{Gray1}
          & s03b  & 95.74 & 97.83 & 93.75 & 91.49 & 93.48 & 86 & 95.74 & 97.83 & 93.75 \\  \rowcolor{Gray1}
          & s04b  & 88.24 & 88.24 & 78.95 & 86.54 & 88.24 & 77.59 & 86.54 & 88.24 & 77.59 \\  \rowcolor{Gray1}
          & s05b  & 98.33 & 98.33 & 96.72 & 98.33 & 98.33 & 96.72 & 95 & 95 & 90.48 \\  \rowcolor{Gray1}
          & s06b  & 100   & 100   & 100   & 100   & 100   & 100   & 100   & 100   & 100 \\  \rowcolor{Gray1}
          & s07b  & 96.55 & 94.92 & 91.8 & 94.83 & 93.22 & 88.71 & 96.55 & 94.92 & 91.8 \\  \rowcolor{Gray1}
          & s08b  & 94.92 & 96.55 & 91.8 & 83.05 & 84.48 & 72.06 & 94.92 & 96.55 & 91.8 \\ 
        \hhline{~----------}  \noalign{\vskip.2pt}
         \rowcolor{Gray1}   
      & & \textbf{93.65} & \textbf{95.33} & \textbf{90.08} & \textbf{91.21} & \textbf{93.07} & \textbf{86.09} & \textbf{93.02} & \textbf{94.92} & \textbf{89.13}  \\  \rowcolor{Gray1} 
 \multirow{-10}{*}{{\rotatebox[origin=c]{90}{\textbf{SBDB}}}}      &    \multirow{-2}{*}{\textbf{Overall}}   & \textbf{$\pm$8.27} & \textbf{$\pm$5.12} & \textbf{$\pm$11.26} & \textbf{$\pm$8.92} & \textbf{$\pm$6.1} & \textbf{$\pm$12.63} & \textbf{$\pm$8.25} & \textbf{$\pm$4.98} & \textbf{$\pm$11.14} \\
    \bottomrule
    \end{tabular}
     \scriptsize
    \vspace{0.1cm}
   Overall performances are given in terms of mean $\pm$ standard deviation.
  \label{tab:DetectResults_systole}%
\end{table}%

\begin{table}
  \caption{Performance of the proposed method in the detection of diastolic fiducial points}
 \scriptsize
  \newcommand{\m}{\hphantom{$-$}}
\newcommand{\cc}[1]{\multicolumn{1}{c}{#1}}
\renewcommand{\tabcolsep}{0.17pc}
\renewcommand{\arraystretch}{1.1}
    \begin{tabular}{ccccccccccc}
    \doubleRule
     & \multirow{2}{*}{\textbf{Record}} & \multicolumn{3}{c}{\textbf{pAC Instant Detection}} & \multicolumn{3}{c}{\textbf{AC Instant Detection}} & \multicolumn{3}{c}{\textbf{MO Instant Detection}} \\
 \cmidrule(lr){3-5}   \cmidrule(lr){6-8}   \cmidrule(lr){9-11}
          &       & \textbf{Se}(\%) & \textbf{+P}(\%) & \textbf{Acc}(\%) & \textbf{Se}(\%) & \textbf{+P}(\%) & \textbf{Acc}(\%) & \textbf{Se}(\%) & \textbf{+P}(\%) & \textbf{Acc}(\%) \\
       \midrule
    \multirow{10}{*}{{\rotatebox[origin=c]{90}{\textbf{NBDB}}}} & s01a  & 98.81 & 100 & 98.81 & 98.21 & 99.7 & 97.92 & 98.21 & 99.4 & 97.63 \\
          & s02a  & 98.9 & 100 & 98.9 & 98.9 & 100 & 98.9 & 98.36 & 99.45 & 97.82 \\
          & s03a  & 99.32 & 99.66 & 98.98 & 98.97 & 99.31 & 98.3 & 99.32 & 99.66 & 98.98 \\
          & s04a  & 99.33 & 100 & 99.33 & 97.64 & 98.31 & 96.03 & 97.98 & 98.64 & 96.68 \\
          & s05a  & 98.82 & 100 & 98.82 & 98.58 & 99.76 & 98.35 & 96.22 & 97.37 & 93.78 \\
          & s06a  & 98.86 & 99.77 & 98.64 & 98.41 & 99.77 & 98.18 & 97.72 & 99.08 & 96.84 \\
          & s07a  & 98.25 & 100 & 98.25 & 95.84 & 97.77 & 93.79 & 95.4 & 97.1 & 92.77 \\
          & s08a  & 100 & 100 & 100 & 99.46 & 99.46 & 98.93 & 90.03 & 90.03 & 81.86 \\ \hhline{~----------}  \noalign{\vskip.2pt}
          & \multirow{2}{*}{\textbf{Overall}} & \textbf{99.04} & \textbf{99.93} & \textbf{98.96} & \textbf{98.25} & \textbf{99.26} & \textbf{97.55} & \textbf{96.65} & \textbf{97.59} & \textbf{94.54} \\
          &       & \textbf{$\pm$0.52} & \textbf{$\pm$0.14} & \textbf{$\pm$0.52} & \textbf{$\pm$1.12} & \textbf{$\pm$0.8} & \textbf{$\pm$1.77} & \textbf{$\pm$2.95} & \textbf{$\pm$3.2} & \textbf{$\pm$5.53} \\
          \hhline{-----------}  \noalign{\vskip.9pt}
          \rowcolor{Gray1}
          & s01b  & 100   & 100   & 100   & 100   & 100   & 100   & 98.15 & 98.15 & 96.36 \\ \rowcolor{Gray1}
          & s02b  & 100   & 89.55 & 89.55 & 100 & 89.55 & 89.55 & 95 & 87.69 & 83.82 \\ \rowcolor{Gray1}
          & s03b  & 100   & 100   & 100   & 100   & 100   & 100   & 97.83 & 97.83 & 95.74 \\ \rowcolor{Gray1}
          & s04b  & 100   & 100   & 100   & 100   & 100   & 100   & 98.08 & 98.08 & 96.23 \\ \rowcolor{Gray1}
          & s05b  & 100   & 100   & 100   & 100   & 100   & 100   & 100   & 100   & 100 \\ \rowcolor{Gray1}
          & s06b  & 98.28 & 100   & 98.28 & 87.72 & 89.29 & 79.37 & 94.74 & 94.74 & 90 \\ \rowcolor{Gray1}
          & s07b  & 100   & 100   & 100   & 96.49 & 96.49 & 93.22 & 100   & 100   & 100 \\ \rowcolor{Gray1}
          & s08b  & 98.33 & 100   & 98.33 & 98.33 & 100   & 98.33 & 98.33 & 100   & 98.33 \\ \hhline{~----------}  \noalign{\vskip.2pt} \rowcolor{Gray1}
          &  & \textbf{99.58} & \textbf{98.69} & \textbf{98.27} & \textbf{97.82} & \textbf{96.92} & \textbf{95.06} & \textbf{97.77} & \textbf{97.06} & \textbf{95.06} \\ \rowcolor{Gray1}
      \multirow{-10}{*}{{\rotatebox[origin=c]{90}{\textbf{SBDB}}}}    &   \multirow{-2}{*}{\textbf{Overall}}    & \textbf{$\pm$0.78} & \textbf{$\pm$3.69} & \textbf{$\pm$3.6} & \textbf{$\pm$4.27} & \textbf{$\pm$4.78} & \textbf{$\pm$7.45} & \textbf{$\pm$1.98} & \textbf{$\pm$4.17} & \textbf{$\pm$5.54} \\
    \bottomrule
    \end{tabular}%
     \scriptsize
    \vspace{0.1cm}
   Overall performances are given in terms of mean $\pm$ standard deviation.
  \label{tab:DetectResults_diastole}%
  \vspace{-0.4cm}
\end{table}%


\subsection{Applicability in Respiratory-Effort Level Identification}

The delineated parameters can be used for detection and classification of many cardiovascular events. 
To show the applicability of the proposed framework, a study of SCG-based cardiorespiratory analysis is performed. 
For this, the effect of respiratory effort levels reflected in SCG-morphology is observed. 
Based on morphological variations, the respiratory conditions of a subject can be categorized as normal breathing and breathlessness conditions. 
A set of twelve extracted features ($\text{f}_k$ $\in \mathbb{R}^{12},~k=1,2,..12 $) is studied for this purpose, and they are AO-AO interval derived heart-rate (HR), relative timing information of AO, IC, AC, pAC and MO with respect to IM, normalized signal amplitudes corresponding to IM, AO, IC, AC, pAC and MO points.   
The features are extracted from each of the cardiac cycles and normalized. In this study, a 40 s SCG-segment from each of the signals is used for feature extraction. The distributions of normalized features are displayed in Fig. \ref{fig:Mean_Features} in terms of mean$\pm$standard-deviation.
More discriminative and independent features are selected after a statistical analysis of feature distributions and the use of standard T-test procedure.
The selected features are $\text{f}_{k^*}$ $\in \mathbb{R}^{8},~k^*=1,2,5,6,7,9,10,11$ for this event classification task.
Finally, support vector machine (SVM) with RBF kernel is used to identify the breathlessness condition.
The choice of training and test cycles is done using the standard 10-fold cross-validation technique \cite{ye2012}.

\begin{figure}[t]\centering
\includegraphics[height=4.5cm]{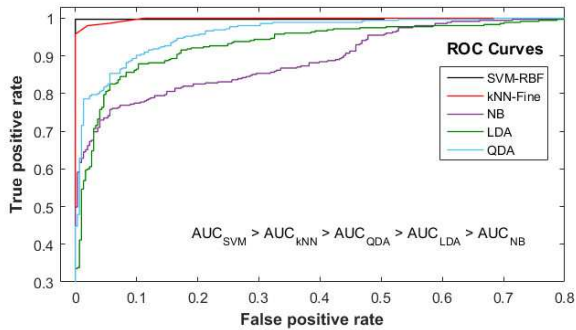}
\caption{ROC of different classifier-models on selected features. SVM-RBF based classifier outperforms others by achieving the largest AUC of 0.9986.}
\label{fig:ROC}
\vspace{-0.2cm}
\end{figure}

The classification performance is evaluated and compared with various other classifiers. 
The performance evaluation is done with three measurement parameters, recognition accuracy $\left(\text{ACC}=\frac{\text{TP+TN}}{\text{TP+TN+FP+FN}}\right)$, true positive rate $\left(\text{TPR}=\frac{\text{TP}}{\text{TP+FN}}\right)$ and false positive rate $\left(\text{FPR}=\frac{\text{FP}}{\text{TN+FP}}\right)$ \cite{tharwat2018}.
The average performance metrics are tabulated in Table \ref{tab:SBclassifyResults}.
The SVM-RBF based classifier outperforms others by producing these metrics as 97.25, 96.72 and 2.29, respectively for considering all features, and 98.35, 97.32 and 0.76, respectively for considering selected feature-set.
Fig. \ref{fig:ROC} shows the estimated receiver operating characteristic (ROC) curves plotted between TPR and FPR, and area under the ROC curve (AUC) for the identification of SB class.
The best performance is achieved with the SVM-RBF classifier. 
This experiment validates that the delineated parameters from the proposed method could be employed not only for the cardiac analysis, but also for the assessment of respiratory system.

\begin{table}
 \caption{Performance results of different classifiers for stopped breathing detection with 10-fold cross validation}
\newcommand{\m}{\hphantom{$-$}}
\newcommand{\cc}[1]{\multicolumn{1}{c}{#1}}
\renewcommand{\tabcolsep}{0.16pc}
\renewcommand{\arraystretch}{1.2}
   \begin{tabular}{cccccccccc}
    \doubleRule 
    \multirow{2}[4]{*}{{\rotatebox[origin=c]{40}{\textbf{Features}}}} & \multirow{2}[4]{*}{{\rotatebox[origin=c]{40}{\textbf{Metrics}}}}  & \multicolumn{2}{c}{\textbf{SVM}}  & \multicolumn{3}{c}{\textbf{kNN}} &  \multirow{2}[4]{*}{\textbf{NB}} 
& \multicolumn{2}{c}{\textbf{Discrim. Ana.}} \\
    \cmidrule(lr){3-4}  \cmidrule(lr){5-7}  \cmidrule(lr){9-10}
          &       & \textbf{RBF} & \textbf{Linear} & \textbf{Fine} & \textbf{Med.} & \textbf{Coarse } &      & \textbf{LDA} & \textbf{QDA} \\
          \midrule
   & {ACC} & \textbf{97.25} & 89.70 & 96.70 & 94.64 & 84.06 & 85.71 & 89.57 & 92.73 \\
     {All}     & {TPR} & \textbf{96.72} & 93.12 & 97.01 & 95.53 & 85.29 & 85.61 & 93.73 & 94.32 \\
         & {FPR} & \textbf{2.29} & 13.22 & 3.54  & 6.08  & 16.99 & 14.22 & 13.97 & 8.63 \\
           \cmidrule(){1-10}
           \rowcolor{Gray1}
    & {ACC} & \textbf{98.35} & 87.23 & 96.02 & 95.61 & 84.90 & 79.40 & 88.19 & 87.65 \\
\rowcolor{Gray1}
    {SF$^\ast$}      & {TPR} & \textbf{97.32} & 89.83 & 95.81 & 95.53 & 87.99 & 75.79 & 92.83 & 91.92 \\
          \rowcolor{Gray1}
         & {FPR} & \textbf{0.76} & 14.99 & 3.81  & 4.31  & 17.76 & 17.51 & 15.74 & 15.98 \\
    \bottomrule
    \end{tabular}%
    \scriptsize
    \vspace{0.1cm}
   Abbreviations -- SVM: support vector machine, RBF: radial basis function, kNN: K-nearest neighbour, Fine: kNN with $K$ = 5, Medium: kNN with $K$ = 11, Coarse: kNN with $K$ = 101, NB: naive Bayes, LDA: linear discriminant analysis based classifier, QDA: quadratic discriminant analysis based classifier.    
    
   $^\ast$SF: selected feature-set. Note that all the metrics are presented here in terms of \%.
  \label{tab:SBclassifyResults}%
\end{table}%

\section{Conclusion}
In this work, a novel framework is proposed for delineation of systole and diastole parameters of a SCG signal with the help of a temporally concurrent fingertip-PPG signal. 
The proposed SCG delineation framework is mainly based on wavelet scalogram, the proposed envelope construction scheme, and the proposed decision rules.
It consists of three major blocks-- estimation of PPG-peaks, estimation of diastole-phases (AC, pAC and MO) in the SCG and estimation of systole-phases (IM, AO and IC) in SCG.
Along with the delineated parameters, LVET and IVRT time-intervals can also be estimated from the parameters.
The experimental results on NBDB and SBDB databases show very promising results, and so, the proposed framework may be deployed for cardiac-cycle event detection for pathological interpretations. 
The salient point of the proposed method is that it can efficiently estimate six important cardiac phases simultaneously, which characterize events associated with both systolic and diastolic regions.
Additionally, as an application for cardiorespiratory analysis, the information extracted from the estimated fiducial points can be used to identify normal breathing and breathlessness conditions. The SVM-RBF classifier gives good results on our selected feature-set.
However, the framework needs to be tested with cardio- and respiratory related pathological scenarios for its clinical applications.



\bibliographystyle{IEEEbib}
\bibliography{tc_References}
\end{document}